\documentclass[11pt]{article}
\newsavebox{\PSLASH}
\sbox{\PSLASH}{$p$\hspace{-1.8mm}/}

\begin{document}
\title{\large \bf  Revisiting the Israel junction conditions in Einstein-Cartan gravity}
\author{ S. Khakshournia$^{1}$
	\footnote{Email Address: skhakshour@aeoi.org.ir} and R.
	Mansouri$^{2,3}$ \footnote{Email Address: mansouri@ipm.ir}\\
	\\
	$^{1}$Nuclear Science and Technology Research Institute (NSTRI), Tehran, Iran\\
	$^{2}$Department of Physics, Sharif University of
	Technology, Tehran, Iran\\
	$^{3}$Institute for Studies in Physics and Mathematics (IPM), Tehran, Iran\\
}\maketitle
\[\]
\[\]

\begin{abstract}
 The Israel junction conditions of a thin shell in the context of Einstein-Cartan gravity are revisited.  It is shown that with a choice of the torsion discontinuity  
taken to be orthogonal to the hypersurface and consistent with the antisymmetric properties of torsion tensor and contorsion tensor,  the generalized asymmetric surface 
energy-momentum tensor turns out to be tangent to the hypersurface while the resulting generalized Israel junction conditions are modified. The main differences to previous works are mentioned.\\

\end{abstract}
\hspace{1.5cm}
\newpage
\section{Introduction}

The Einstein-Cartan (EC) theory of gravitation is a natural extension of general relativity that accounts for the presence of spacetime torsion which could be arising from the spinning properties of the matter distribution. In this context, dynamics of the spacetime is determined by torsion being triggered by the intrinsic angular momentum of the matter, and the curvature being stemmed from the mere presence of matter (see \cite{blagojevic2013gauge} for a recent review and references therein).\\
The first extension of the junction conditions to the EC gravity was made in \cite{arkuszewski1975matching} presented in the language of differential forms. The paper is, however, restricted to the case of boundary surfaces in general relativity where the extrinsic curvature is continuous across the hypersurface. Later, Bressange \cite{bressange2000extension} extended the unified description of thin shells of any type, including the null case, provided by Barrabes-Israel \cite{barrabes1991thin} in the presence of torsion. In his approach the contorsion discontinuity, naturally appearing in the expression for the shell's energy-momentum tensor, is taken to be orthogonal to the hypersurface, yielding a modification to the surface energy-momentum tensor and the resulting junction conditions. According to this modification, the Israel junction conditions turn out to be formally the same as in general relativity \cite{Israel1966NuCimen}, but the tensor representing the jump of the normal derivatives of the metric across the nonlightlike shell is replaced with a non-symmetric one splitting into a Riemann part and a Cartan part. Recently, based on the Bressange's approach, the junction conditions of two generic spacetimes through a non-lightlike hypersurface in the context of $ f(R) $ gravity with torsion have been derived via the definition of a generalized extrinsic curvature tensor which splits into a Riemann part and a Cartan one \cite{vignolo2018junction}.\\
On the other hand,  Hoff et al. presented an extension of junction conditions for a timelike hypersurface to cover the EC gravity in the context of Braneworld scenarios \cite{da2009braneworld}(see also \cite{da2010possible}). They assumed the torsion discontinuity to be orthogonal to $ \Sigma $, without considering the antisymmetric property of the torsion tensor on the last two indices. As a result, there appears an extra term, associated with the torsion sector of the connection, in the shell energy-momentum tensor. Requiring the tensor be tangent to the hypersurface leads to the vanishing of that term, and makes the Israel junction conditions unaltered in the presence of torsion.\\
There is a third approach to the subject adopted by Maier et al. \cite{maier2011brane} in the context of braneworld models with $ Z_{2} $ symmetry. They assumed the torsion in the bulk to be continuous across the brane just as the metric tensor but its first derivatives which appear in the effective field equations on the 3-brane can be discontinuous. In their work, the Israel junction conditions are extended such that the extrinsic curvature tensor splits into a symmetric part connecting to the matter distribution on the brane and an antisymmetric part including the 5-dimension torsion tensor projected onto the brane. But this splitting of the extrinsic curvature tensor  is confined to the highly symmetric spacetimes in which the energy-momentum tensor retains its familiar symmetry despite the presence of torsion.\\
This paper aims to apply a proper choice for the torsion discontinuity taken to be orthogonal to a timelike or spacelike hypersurface to find the generalized form of the Israel junction conditions in the presence of torsion. Very recently this approach was applied for a null shell in the context of Einstein-Cartan theory of gravity \cite{khakshournia2019CQG}.\\ 
\textit{Conventions.}  Natural geometrized units, $G=c=1$,  are used throughout the paper. The hypersurface is denoted by $\Sigma$. We use square brackets [F] to denote the jump of any quantity F across $\Sigma$. Latin indices range over the intrinsic coordinates 
of $\Sigma$ denoted by $\xi^{a}$, and Greek indices over the coordinates of the 4-manifolds. As we are going to work with distributional valued tensors, there may be terms is a tensor quantity $F$ proportional to some $\delta$-function distribution. These terms are indicated by $\breve{F}$.\\

   \section{The Einstein-Cartan gravity}
   In a non-Riemannian manifold with torsion, the torsion tensor is defined by 
   \begin{equation}\label{tortdef}
   	{T^{\sigma}}_{\mu\nu}={\Gamma^{\sigma}}_{\nu\mu}
   	-{\Gamma^{\sigma}}_{\mu\nu},
   \end{equation}
   with $ {T^{\sigma}}_{\mu\nu}=-{T^{\sigma}}_{\nu\mu} $. Imposing the metricity condition $ \nabla_{\sigma}g_{\mu\nu}=0 $, the asymmetric connection can be decomposed in a unique way as 
   \begin{equation}\label{tordecom}
   	{\Gamma^{\sigma}}_{\mu\nu}= \mathring{{\Gamma}^{\sigma}}_{\mu\nu}
   	+{K^{\sigma}}_{\mu\nu},
   \end{equation}
 where $ \mathring{{\Gamma}^{\sigma}}_{\mu\nu} $ denotes the Christoffel symbols
 \begin{equation}\label{tordchrist}
 \mathring{{\Gamma}^{\sigma}}_{\mu\nu}= \frac{1}{2}g^{\sigma\lambda}\left( \partial_{\mu}g_{\nu\lambda}+\partial_{\nu}g_{\mu\lambda}-\partial_{\lambda}g_{\mu\nu}\right), 
\end{equation}
  and $ {K^{\sigma}}_{\mu\nu} $ are the components of the contorsion 
 of the connection related to the torsion by
   \begin{equation}\label{torcotdef}
   	K_{\sigma\mu\nu}=\frac{1}{2}(T_{\mu\sigma\nu}
   	+T_{\nu\sigma\mu} -T_{\sigma\mu\nu}),
   \end{equation}
   with $ K_{\sigma\mu\nu}=-K_{\mu\sigma\nu} $. 
The Einstein-Cartan field equations are written as
   \begin{equation}\label{torfieldeq1}
   	G_{\mu\nu}=8\pi T_{\mu\nu},
   \end{equation}
   \begin{equation}\label{torfieldeq2}
   	{T^{\mu}}_{\nu\sigma}+ \delta^{\mu}_{\nu} {T^{\rho}}_{\sigma\rho}- \delta^{\mu}_{\sigma} {T^{\rho}}_{\nu\rho}=8\pi {S^{\mu}}_{\nu\sigma},
   \end{equation}
   where $ G_{\mu\nu} $ is the non-symmetric Einstein tensor constructed out of the Ricci tensor $ R_{\mu\nu} $ based on the non-symmetric connection ${\Gamma^{\sigma}}_{\mu\nu} $, $ T_{\mu\nu}$  is the non-symmetric energy-momentum tensor of the matter distribution in the spacetime and $ {S^{\mu}}_{\nu\sigma} $ is the spin tensor representing the density of the intrinsic angular momentum (spin) in the matter related to the torsion tensor 
in a purely algebraic way according to Eq. (\ref{torfieldeq2}). \\
   
    \section{The thin shell formalism in the presence of torsion}
  
    Consider $x^{\mu}$ to be an admissible coordinate system in a coordinate neighborhood that includes the non-null hypersurface $\Sigma$ extending into both 
non-Riemannian spacetimes $\cal M^{\pm}$. Let  $\Phi (x^{\mu})=0$ be the local equation of  $\Sigma$ in this coordinate system. The domains in which $\Phi$ is positive or negative are contained in $\cal M^{+}$ or $\cal M^{-}$, respectively. One can write the metric in the domain of admissible coordinate system as a distribution-valued tensor
     \begin{equation}\label{distmetric}
        g_{\mu\nu}=	g^{+}_{\mu\nu}\Theta(\Phi)+	g^{-}_{\mu\nu}\Theta(-\Phi),
    \end{equation}
 where $ \Theta(\Phi) $ is the step function and $ g^{-}_{\mu\nu} $ and $ g^{+}_{\mu\nu} $ are metrics in $ \cal M^{-} $ and $ \cal M^{+} $, continuously 
glued at $ \Sigma $. The corresponding tangent vectors on $ \Sigma $ are $ e_{(a)}=\partial/\partial\xi^{a} $, and the normal vector defined 
by $n_{\mu}=\epsilon|\alpha|^{-1}\partial_{\mu}\Phi$, with $\alpha$ being the normalizing factor such that $  n_{\mu}n^{\mu}=\epsilon$, and $ \epsilon=+1(-1)$ if the hypersurface is timelike (spacelike). The associated jumps on $\Sigma$ expressed in the admissible coordinates $x^{\mu}$ should vanish:
    $[g_{\mu\nu}]=[n^{\mu}]=[e^{\mu}_{a}]=[\epsilon]=[\alpha]=0$.  Assuming the induced metric $ h_{ab}=e_{(a)}.e_{(b)} $ on $ \Sigma $,  the same on both sides of the hypersurface, the completeness relations for the basis are given by 
     \begin{equation}\label{completerel}
            g^{\mu\nu}=h^{ab}e^{\mu}_{a}e^{\nu}_{b}+\epsilon n^{\mu}n^{\nu},
      \end{equation}
   where $ h^{ab} $ is the inverse of the induced metric. Besides,  the purely tangential part of the torsion tensor is assumed to be continuous across $ \Sigma $ \cite{bressange2000extension}:
       \begin{equation}\label{tortancomt}
       	e^{\mu}_{a}e^{\nu}_{b}e^{\sigma}_{c}[T_{\mu\nu\sigma}]=[T_{abc}]=0. 
       \end{equation}
   The metric continuity condition $[g_{\mu\nu}]=0 $ in the admissible coordinates $ x^{\mu} $ guarantees that the tangential derivatives of the metric are continuous: $[g_{\mu\nu,\sigma}]e^{\sigma}_{a}=0$. This means that if however, the derivative of metric is to be discontinuous, this discontinuity must be directed along the normal vector $ n^{\alpha} $ \cite{poisson2004Rel}
   \begin{equation}\label{metricdisc}
     [g_{\mu\nu,\sigma}] =\gamma_{\mu\nu}n_{\sigma},	 
   \end{equation}
   where the tensor $\gamma_{\mu\nu}$ is given explicitly by $ \gamma_{\mu\nu}= \epsilon[g_{\mu\nu,\sigma}]n^{\sigma} $. Now, the discontinuity of the Christoffel symbols across $ \Sigma $ is given by    
      \begin{equation}\label{Chrisdisc}
      	[{\stackrel{\circ}{\Gamma}}{}^\sigma_{\;\,\mu\nu}] =
      	\frac{1}{2}(\gamma^\sigma_{\;\,\mu}
      	n_\nu+\gamma^\sigma_{\;\,\nu} n_\nu - \gamma_{\mu\nu}n^\sigma).
      \end{equation}
Taking into account the continuity of tangential part of the torsion tensor as expressed in (\ref{tortancomt}), any possible discontinuity must be normal 
to $ \Sigma $. Hence, the discontinuity of the torsion tensor, as a primary geometric object, across $ \Sigma $ can be decomposed as
      \begin{equation}\label{tordisc}
       [{T^{\sigma}}_{\mu\nu}]=\zeta_{\mu\nu}n^{\sigma},
      \end{equation}
for some  tensor $ \zeta_{\mu\nu} $ given explicitly by $ \zeta_{\mu\nu}=\epsilon n_{\sigma}[{T^{\sigma}}_{\mu\nu}] $. Consistency with the antisymmetric property 
of ${ T^{\sigma}}_{\mu\nu} $ on the last two indices requires the tensor $ \zeta_{\mu\nu} $  be antisymmetric. Note the difference between the choice of
(\ref{tordisc}) and the choice of the torsion tensor discontinuity orthogonal to $ \Sigma $ as $ [{T^{\sigma}}_{\mu\nu}]=\zeta_{\mu}^{\;\,\sigma}n_{\nu} $ 
made in \cite{da2009braneworld} without considering the antisymmetric property of the torsion tensor on the last two indices. Now using Eq. (\ref{torcotdef}), the 
contorsion discontinuity across $ \Sigma $ is written as
   \begin{eqnarray}\label{cotdis1}
           	[K^\sigma_{\;\,\mu\nu}] &=&
           \frac{1}{2}(\zeta^{\sigma}_{\;\,\nu} n_\mu+\zeta^{\sigma}_{\;\,\mu} n_\nu
           	 -\zeta_{\mu\nu} n^{\sigma}).  
   \end{eqnarray}
 This contorsion discontinuity induced by our choice of the torsion discontinuity (\ref{tordisc}) is automatically consistent with the antisymmetric 
property of the contorsion tensor on its first two indices. This result, obtained from the choice (\ref{tordisc}), should be compared with the 
choice $[{K^{\sigma}}_{\mu\nu}]=\frac{1}{2}\beta_{\mu\nu}n^{\sigma} $, for some tensor $ \beta_{\mu\nu} $, assumed by Bressange \cite{bressange2000extension}. Substitution of
the Eqs. (\ref{Chrisdisc}) and (\ref{cotdis1}) into  Eq. (\ref{tordecom}) leads to the following expression for the discontinuity of the non-symmetric connection across $ \Sigma$   
  \begin{eqnarray}\label{asymconjump}
   	[\Gamma^\sigma_{\;\,\mu\nu}] &=&
  	\frac{1}{2}(\gamma^\sigma_{\;\,\mu}
   	n_\nu+\gamma^\sigma_{\;\,\nu} n_\mu - \gamma_{\mu\nu}n^\sigma)+\frac{1}{2}(\zeta^{\sigma}_{\;\,\nu} n_\mu+\zeta^{\sigma}_{\;\,\mu} n_\nu	-\zeta_{\mu\nu} n^{\sigma}).
  \end{eqnarray}
Writing the distribution-valued Riemann tensor as 
   $ R^{\alpha}_{\;\,\mu\sigma\nu}={R^{+\alpha}}_{\mu\sigma\nu}\Theta(\Phi) +{R^{-\alpha}}_{\mu\sigma\nu}\Theta(-\Phi)
   +\breve{R}^{\alpha}_{\;\,\mu\sigma\nu}\delta(\Phi)$, where $ \breve{R}^{\alpha}_{\;\,\mu\sigma\nu}=\epsilon([{\Gamma^{\alpha}}_{\mu\nu}]n_{\sigma}-
   [\Gamma^{\alpha}_{\;\,\mu\sigma}]n_{\nu})$ \cite{poisson2004Rel},  we can obtain the singular part of the Riemann tensor using Eq. (\ref{asymconjump})
    \begin{eqnarray}\label{Riemtensing} 
      	\breve{R}^{\alpha}_{\;\,\mu\sigma\nu}&=&-\left.\frac{\epsilon}{2}(\gamma^{\alpha}_{\;\,\sigma}	n_\mu n_{\nu} -\gamma_{\nu}^{\;\,\alpha}n_\mu n_{\sigma} +
      	\gamma_{\mu\nu} n^\alpha n_\sigma-\gamma_{\mu\sigma} n^\alpha n_\nu)\right. \nonumber \\
      	&+&\frac{\epsilon}{2}(\zeta^{\alpha}_{\;\,\nu} n_\mu n_{\sigma}-\zeta^{\alpha}_{\;\,\sigma} n_\mu n_{\nu}-\zeta_{\mu\nu} n^{\alpha}n_{\sigma}+\zeta_{\mu\sigma} n^{\alpha}n_{\nu}).
     \end{eqnarray}
 Contractions of Eq. (\ref{Riemtensing}) yields the singular part $ \breve{R}_ {\mu\nu} $ of the Ricci tensor  
   \begin{eqnarray}\label{Ricctsing}
   	\breve{R}_{\mu\nu}&=&-\left.\frac{\epsilon}{2}(\gamma n_\mu n_\nu-\gamma_{\nu\sigma}
   n^\sigma	n_\mu-\gamma_{\mu\sigma} n^\sigma n_\nu+\epsilon\gamma_{\mu\nu})\right. +\frac{\epsilon}{2}(\zeta_{\mu\sigma} n_\nu n^{\sigma}-\zeta_{\nu\sigma} n_\mu n^{\sigma}-\epsilon\zeta_{\mu\nu}),
    \end{eqnarray}
 and the singular part $ \breve{R} $ of the Ricci scalar is determined correspondingly
   \begin{eqnarray}\label{Riccssing}
   	\breve{R} =
   	g^{\mu\nu}\breve{R}_{\mu\nu} = -\epsilon(\epsilon\gamma-\gamma_{\rho\sigma} n^\rho
   	n^\sigma).
   \end{eqnarray}
 Finally, we arrive at  the following form  for the singular part $ \breve{G}_{\mu\nu} $ of the Einstein tensor 
   \begin{eqnarray}\label{Eintentsing}
   	\breve{G}_{\mu\nu}&=&-\frac{\epsilon}{2}(\gamma n_{\mu}n_{\nu}+g_{\mu\nu}\gamma_{\rho\sigma}n^{\rho}n^{\sigma}-\gamma_{\nu\sigma}
   	n^\sigma n_\mu-\gamma_{\mu\sigma}n^{\sigma}n_{\nu}+\epsilon\gamma_{\mu\nu}-\epsilon\gamma g_{\mu\nu})\\ \nonumber
   &+& \frac{\epsilon}{2}(\zeta_{\mu\sigma} n_\nu n^{\sigma}-\zeta_{\nu\sigma} n_\mu n^{\sigma}-\epsilon\zeta_{\mu\nu}).
   	\end{eqnarray}
   	Writing the distribution-valued energy-momentum tensor as
   	 \begin{equation}\label{diststrsengten}
   	 	T_{\mu\nu}=	T^{+}_{\mu\nu}\Theta(\Phi)+	T^{-}_{\mu\nu}\Theta(-\Phi)+S_{\mu\nu}\delta(\Phi),
   	 \end{equation}
    and noting $\breve{G}_{\mu\nu}=8\pi S_{\mu\nu}$, then the surface energy-momentum tensor of the shell $S_{\mu\nu}  $ is found to be 
   \begin{eqnarray}\label{surfaceengten}
   	8\pi\epsilon S_{\mu\nu}&=&-\frac{1}{2}(\gamma n_{\mu}n_{\nu}+g_{\mu\nu}\gamma_{\rho\sigma}n^{\rho}n^{\sigma}-\gamma_{\nu\sigma}n^\sigma n_\mu-\gamma_{\mu\sigma}n^{\sigma}n_{\nu}+\epsilon\gamma_{\mu\nu}-\epsilon\gamma g_{\mu\nu})\\ \nonumber
   	  &+& \frac{1}{2}(\zeta_{\mu\sigma} n_\nu n^{\sigma}-\zeta_{\nu\sigma} n_\mu n^{\sigma}-\epsilon\zeta_{\mu\nu}). 	
   \end{eqnarray}
 It is easy to see that by virtue of the antisymmetric property of $  \zeta_{\mu\nu}$ the following two conditions hold:
   \begin{equation}\label{tangenttensor}
   (i)S_{\mu\nu}n^{\mu}=0,	 \hspace{1cm}  (ii)S_{\mu\nu}n^{\nu}=0.
   \end{equation}
  These are the necessary and sufficient conditions for the asymmetric tensor $ S^{\mu\nu} $ to be purely tangent to $ \Sigma $. We, therefore see that there is no need to impose any additional constraints to ensure that the surface tensor $ S_{\mu\nu} $ is indeed tangent to $ \Sigma $, in contrast to \cite{bressange2000extension,da2009braneworld}. 
 Based on the conditions (\ref{tangenttensor}), the following decomposition can be made: 
   \begin{equation}\label{projecttensor}
     S^{\mu\nu}=S^{ab}e^{\mu}_{a}e^{\nu}_{b}.
     \end{equation}
     Now, $ S_{ab}=S_{\mu\nu}e^{\mu}_{a}e^{\nu}_{b} $ is an asymmetric three-tensor to be evaluated as
  \begin{eqnarray}\label{projecttensor2}
     16\pi S_{ab}&=& -\epsilon\gamma_{\rho\sigma}n^{\rho}n^{\sigma}h_{ab}+\gamma h_{ab}-\gamma_{ab}-\zeta_{ab},\nonumber\\
      &=&-\gamma_{\rho\sigma}(g^{\rho\sigma}-e^{\rho}_{c}e^{\sigma}_{d}h^{cd})h_{ab}+\gamma h_{ab}-\gamma_{ab}-\zeta_{ab},\nonumber\\ &=&-(\gamma_{ab}
      +\zeta_{ab})+\gamma_{\rho\sigma}e^{\rho}_{c}e^{\sigma}_{d}h^{cd}h_{ab},\nonumber\\
      &=&-(\gamma_{ab}+\zeta_{ab})+\gamma h_{ab},
   \end{eqnarray}
   where the completeness relations (\ref{completerel}) have been used. On the other hand, an asymmetric extrinsic curvature tensor $ {\cal K}_{ab} $ may now be introduced with the jump across $ \Sigma $ given by
    \begin{eqnarray}\label{extrintensjum1}
   [{\cal K}_{ab}]&=&[\nabla_{\mu}n_\nu]e^{\mu}_{a}e^{\nu}_{b},\nonumber\\
   &=&-[\Gamma^\sigma_{\;\,\mu\nu}]n_{\sigma}e^{\mu}_{a}e^{\nu}_{b},\nonumber\\
    &=&\frac{\epsilon}{2}(\gamma_{ab}+\zeta_{ab}),
    \end{eqnarray}
    where Eq. (\ref{asymconjump}) has been used. Using this equation for the jump of the extrinsic curvature, and the final form of Eq. (\ref{projecttensor2}), we
    end up to the following form for the generalized Israel junction conditions in the presence of torsion:
   \begin{equation}\label{dijc}
   [{\cal K}_{ab}]-[{\cal K}]h_{ab}=-8\pi\epsilon S_{ab},
   \end{equation}
 with $ [{\cal K}]=\gamma  $, since due to the antisymmetric property of $ \zeta_{\mu\nu} $, one gets $ \zeta=0 $. Therefore, it is seen that in the context of Einstein-Cartan gravity, the junction conditions take the same form (\ref{dijc}) as 
in general relativity with $  [{\cal K}_{ab}] $ being now a nonsymmetric tensor splitting into a Riemann part $ \gamma_{ab} $ and a Cartan part $ \zeta_{ab} $. The authors in
 \cite{da2009braneworld}, had to assume the vanishing of a torsion term in the obtained shell energy-momentum tensor to ensure that it is tangent to $ \Sigma $, resulting in the Israel junction conditions without any modification. The need for this extra assumption made in \cite{da2009braneworld} is due to the improper choice of the decomposition of the torsion tensor discontinuity.\\     
 \section{Conclusion}
We have extended the Israel junction conditions to the Einstein-Cartan theory of gravity, based on the assumption of the torsion discontinuity to be orthogonal to the hypersurface. The resulting Israel junction conditions in the presence of torsion are then modified such that the jump of the extrinsic curvature across the hypersurface turns out to be a nonsymmetric tensor splitting into a Riemann part and a Cartan part. The unique choice of (\ref{tordisc}) for the discontinuity of the torsion tensor across the hypersurface is not only consistent with the antisymmetric properties of torsion and contorsion tensors but also results in a surface energy-momentum tensor which is automatically tangent to the hypersurface.\\

\end{document}